\documentclass[preprint,showkeys]{revtex4-2}

\usepackage{graphicx}
\usepackage{dcolumn}
\usepackage{bm}
\usepackage{amsfonts}
\usepackage{amsmath}
\usepackage{braket}
\usepackage[utf8]{inputenc}
\usepackage{hyperref}
\usepackage{color,soul}

\begin{document}

\title{Eigenmode-Guided Amplification via Spatiotemporal Active Acoustic Metamaterials}

\author{Wai Chun Wong$^{1}$}
\author{Gregory Chaplain$^{2}$}
\author{Jensen Li$^{1}$}
\email{j.li13@exeter.ac.uk}

\affiliation{$^{1}$Centre for Metamaterial Research and Innovation, Department of Engineering, University of Exeter, United Kingdom.}
\affiliation{$^{2}$Centre for Metamaterial Research and Innovation, Department of Physics and Astronomy, University of Exeter, United Kingdom.}

\begin{abstract}
We present a spatiotemporal gain–loss framework for eigenmode steering in coupled acoustic resonators.  
A cross-coupled gain–loss coefficient links the gain of one resonator to the intensity of its partner, creating nonlinear feedback that conserves total energy while driving the system toward the eigenmode associated with the eigenvalue having the largest imaginary part—a deterministic eigenmode collapse.  
Spatial gain–loss profiles shape the eigenvalue spectrum and attractor landscape, while temporal modulation governs the transition dynamics.  
When symmetry prevents direct access to a target eigenmode, controlled spatiotemporal perturbations enable otherwise symmetry-forbidden transitions and accelerate convergence.  
Within this framework, parity–time ($\mathcal{PT}$) symmetry appears as a special case, allowing tunable switching between collapse and Rabi-like oscillations near the exceptional point.  
Full-wave simulations of coupled Helmholtz resonators confirm precise and programmable acoustic energy routing, establishing spatiotemporal gain–loss engineering as a route to reconfigurable wave control and analog information processing.
\end{abstract}

\maketitle

\section{Introduction}

Time-varying media are systems whose physical parameters are deliberately modulated in time to control how waves evolve, enabling transitions between different physical states or modes.  
When the Hamiltonian varies slowly within the adiabatic regime, the system remains in the instantaneous eigenstate of the Hamiltonian~\cite{Born1928}, allowing stable, well-controlled mode evolution.  
When the modulation rate increases, however, the evolution departs from the adiabatic path, giving rise to nonadiabatic transitions such as Landau--Zener tunneling~\cite{Wittig2005} in two-level systems, where the transition probability depends sensitively on the temporal sweep rate through an avoided crossing.  
Recent advances in metamaterials~\cite{Yu2011,Zheng2015,Soukoulis2007,Pendry2000,Wang2018,Ni2015,Zhang2008,Li2019,Zhou2020,Li2013,Li2004,Wu2014,Popa2018,Xie2014,Yan2018,Zhu2014} now allow temporal modulations comparable to the wave oscillation period, opening access to new dynamic regimes of wave manipulation.  
Such systems have revealed rich phenomena including nonreciprocal propagation~\cite{Guo2019}, enhanced nonlinearities~\cite{Taravati2018,Estep2014,Tirole2024}, broadband frequency conversion~\cite{Moussa2023}, and temporal analogues of spatial interference such as diffraction in time~\cite{Tirole2023}.  
They also provide new routes for parametric amplification and time-dependent control of energy exchange~\cite{Galiffi2019,Pendry2021,Yang2023,Koutserimpas2018}.

In parallel, non-Hermitian physics—particularly parity–time ($\mathcal{PT}$) symmetry—has revealed how spatially balanced gain and loss can reshape modal spectra and direct energy flow.  
$\mathcal{PT}$-symmetric systems have demonstrated selective amplification and mode conversion~\cite{Bender1998,ElGanainy2018,Feng2014,Hodaei2016,Teimourpour2017,Zhong2020,Wen2023}, showing that the spatial gain–loss distribution governs competition between modes.  
However, most realizations remain static in time, limiting their ability to control when and how transitions between eigenmodes occur.  
Embedding $\mathcal{PT}$ symmetry into a time-varying framework offers a new degree of freedom: spatial gain–loss contrast defines the stability and rate of amplification, while temporal modulation dictates the trajectory of modal evolution.

Here we introduce a spatiotemporal gain–loss framework that achieves \emph{eigenmode steering} in coupled acoustic Helmholtz resonators.  
The total acoustic energy remains conserved through a \emph{cross-coupled gain–loss coefficient} that links the gain of one resonator to the intensity of its partner, producing nonlinear feedback that drives the system toward fixed-point solutions corresponding to the eigenmodes of an effective non-Hermitian Hamiltonian.  
The eigenmode associated with the eigenvalue having the largest imaginary part naturally emerges as the attractor, realizing a deterministic form of \emph{eigenmode collapse}.  
By tailoring the spatial distribution of gain and loss, the eigenvalue spectrum and attractor landscape can be engineered, while temporal modulation governs transitions between eigenmodes.  
We further show that introducing controlled \emph{spatiotemporal perturbations} enables transitions that would otherwise be symmetry-forbidden, providing an additional lever for accelerating convergence and controlling modal pathways.  
Through a $\mathcal{PT}$-symmetric implementation, we demonstrate controlled switching between collapse and Rabi-like oscillations, and programmable cyclic energy routing across multiple resonators.  
This establishes spatiotemporal gain–loss engineering as a general framework for reconfigurable acoustic manipulation, time-varying metamaterials, and analog information processing.

\section{Results and Discussions}

\subsection{Dimer Model for Eigenmode Steering}

Resonant systems support discrete eigenmodes whose amplitudes evolve under the combined effects of coupling and dissipation.  
To expose the basic mechanism of \emph{eigenmode steering}, we begin with the simplest case—a dimer of two coupled resonators—in which the gain–loss distribution is programmed to guide the system toward a target eigenmode.  
We cast the coupled-resonator dynamics in Hamiltonian form to unify coupling, gain, and loss, and to reveal the nonlinear feedback that governs the evolution.  
The resulting Hamiltonian is nonlinear because the effective gain and loss at each site depend on the instantaneous field amplitudes, producing an intensity-dependent feedback.

The dynamics are described by the coupled-mode equation
\begin{equation}
    i\frac{d}{dt}
    \begin{pmatrix}
        \psi_a \\[4pt]
        \psi_b
    \end{pmatrix}
    = H
    \begin{pmatrix}
        \psi_a \\[4pt]
        \psi_b
    \end{pmatrix},
    \label{eq:dimer_eq}
\end{equation}
with the nonlinear Hamiltonian
\begin{equation}
    H = \frac{1}{2}
    \begin{pmatrix}
        i g \frac{|\psi_b|^2}{E} & -\frac{\kappa}{2} \\[6pt]
        -\frac{\kappa}{2} & - i g \frac{|\psi_a|^2}{E}
    \end{pmatrix},
    \label{eq:Hamiltonian}
\end{equation}
where $\psi_a$ and $\psi_b$ are the complex envelopes of modes $a$ and $b$, $\kappa$ is the coupling coefficient, and $g$ is the \emph{cross-coupled gain--loss coefficient}.  
This coefficient implements a nonlinear inter-site feedback: the gain (loss) experienced by one resonator depends on the intensity of its coupled partner.  
For $g>0$, mode $a$ experiences gain $\propto |\psi_b|^2$, while mode $b$ experiences loss $\propto |\psi_a|^2$.  
Although these terms render the system non-Hermitian, their cross-coupled form ensures that the total energy
\[
E = |\psi_a|^2 + |\psi_b|^2
\]
remains a constant of motion (see Supplementary Information for derivation and extension to non-identical resonators).  
This nonlinearity provides a natural saturation mechanism that prevents unbounded growth or decay, enabling sustainable eigenmode steering.  
The normalization by $E$ in Eq.~(\ref{eq:Hamiltonian}) also guarantees scale invariance.

\begin{figure}[h]
    \centering
    \includegraphics[width=0.6\linewidth]{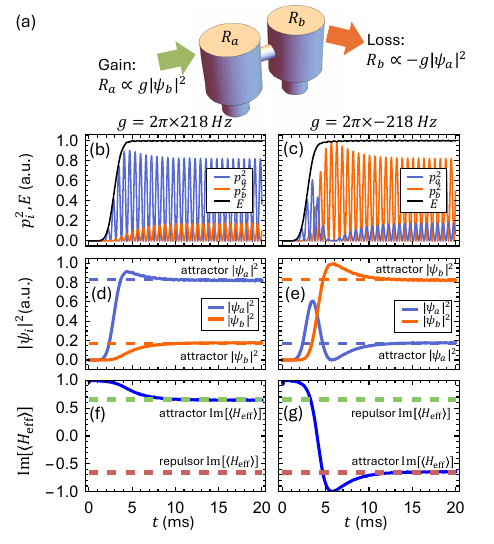}
    \caption{
    (a) Schematic of the nonlinear acoustic dimer composed of two coupled Helmholtz resonators, where the \emph{cross-coupled gain--loss coefficient} modulates wall resistance via $R_{a,b} \propto \mp g|\psi_{b,a}|^2$. 
    (b, c) Simulated squared pressures $p_i^2$ and total energy $E$, and (d, e) squared envelope amplitudes $|\psi_i|^2$, for (b, d) $g = 2\pi \times 218~\mathrm{Hz}$ and (c, e) $g = -2\pi \times 218~\mathrm{Hz}$. 
    Positive $g$ concentrates energy in resonator~$a$, while negative $g$ favors resonator~$b$, consistent with the attractor eigenmodes predicted by the effective Hamiltonian. 
    Dashed horizontal lines denote steady-state fixed points. 
    (f, g) Projection metric $\mathrm{Im}[\langle \psi | H_\mathrm{eff} | \psi \rangle]$ for the simulated field (solid) and the attractor/repulsor eigenvectors of $H_\mathrm{eff}$ (dashed), confirming convergence to the corresponding eigenmode after the transient.
    }
    \label{fig:fig1}
\end{figure}

Such a nonlinear dimer can be implemented acoustically using two coupled Helmholtz resonators operating near a carrier frequency, e.g. chosen as $\omega/2\pi \approx 800~\mathrm{Hz}$. 
Since the pressure fields oscillate predominantly at this frequency, each resonator is represented by a slowly varying complex envelope $\psi_{a,b}$ extracted from the analytic pressure signal
\[
\psi_{a,b} e^{-i\omega t} = p_{a,b} + i\,\mathcal{H}[p_{a,b}],
\]
where $p_{a,b}$ are the measured pressures and $\mathcal{H}[p_{a,b}]$ their Hilbert transforms, approximated under the narrowband condition as $\mathcal{H}[p_{a,b}] \approx -\omega\!\int p_{a,b}\,dt$. 
This envelope description connects the data to Eq.~(\ref{eq:dimer_eq}) and justifies removing common diagonal terms in $H$, since the model now governs slowly varying envelopes rather than rapidly oscillating carriers.

Inter-resonator coupling $\kappa$ arises from the connecting bridge (hollow tube), while the cross-coupled gain–loss action is introduced through time-varying wall resistances $R_{a,b}$ at the upper cavity boundaries, following $R_{a,b} \propto \mp g|\psi_{b,a}|^2$ (Fig.~\ref{fig:fig1}(a)). 
Positive and negative resistances correspond to acoustic loss (orange) and gain (green), implemented numerically via impedance boundary conditions or experimentally via feedback loops linking microphones and speakers on the cavity walls~\cite{Shen2019,Rupin2019}. 
Two-dimensional finite-element simulations (COMSOL Multiphysics) determine $\kappa = 2\pi \times 163~\mathrm{Hz}$ and $g = \pm 2\pi \times 218~\mathrm{Hz}$; geometric details are provided in the Supplementary Information.

Figures~\ref{fig:fig1}(b)–(e) present the simulated squared pressures $p_{a,b}^2$ and corresponding complex evelope amplitudes $|\psi_{a,b}|^2$ for $g = \pm 2\pi \times 218~\mathrm{Hz}$. 
Within approximately $5~\mathrm{ms}$ after energy injection through the neck of resonator~$a$, the system evolves toward distinct attractor states: for $g>0$ energy concentrates in resonator~$a$, whereas for $g<0$ it localizes in resonator~$b$. 
The total energy $E = |\psi_a|^2 + |\psi_b|^2$, shown in black, remains constant after the transient, confirming that the cross-coupled gain–loss feedback provides a saturation mechanism that sustains stable eigenmode steering without the exponential divergence typical of conventional gain systems.

The sustained convergence in Fig.~\ref{fig:fig1} shows that the system evolves toward a unique modal state—\emph{eigenmode collapse}. 
By analogy with quantum wavefunction collapse (but here fully deterministic), the final state coincides with an eigenmode of a Hamiltonian selected by the gain–loss configuration. 
To describe this process, we introduce a linear \emph{effective Hamiltonian} capturing the dominant modal dynamics:
\begin{equation}
    H_{\mathrm{eff}} = \frac{1}{2}
    \begin{pmatrix}
        i\frac{g}{2} & -\frac{\kappa}{2} \\[6pt]
        -\frac{\kappa}{2} &  - i\frac{g}{2}
    \end{pmatrix}.
    \label{eq:Heff}
\end{equation}
The nonlinear Hamiltonian in Eq.~(\ref{eq:Hamiltonian}) decomposes as 
$H = H_{\mathrm{eff}} - i G I_2$, 
where $G = g (|\psi_a|^2 - |\psi_b|^2)/(4E)$ is a common nonlinear gain–loss term and $I_2$ is the $2\times2$ identity. 
Thus $H_{\mathrm{eff}}$ fixes the modal structure and eigenvalue spectrum, while $G I_2$ regulates the overall amplitude to conserve total energy. 

Working in the eigenbasis of $H_\mathrm{eff}$, writing $(\psi_a,\psi_b)=a_1\psi_1+a_2\psi_2$ with eigenvalues $\lambda_{1,2}$, we obtain
\begin{equation}
    \frac{d}{dt}\ln\!\left(\frac{a_1}{a_2}\right)=-\,i(\lambda_1-\lambda_2),
    \label{eq:ratio_dynamics}
\end{equation}
so that $a_1/a_2\propto e^{-i(\lambda_1-\lambda_2)t}$. 
The magnitude ratio grows or decays exponentially with rate $\mathrm{Im}[\lambda_1-\lambda_2]$,  
driving the system toward the eigenmode with the larger imaginary part of eigenvalue (the \emph{attractor}), while the other acts as a \emph{repulsor}. 
For the present parameters, the attractor eigenmodes are $\psi_+ = \{0.91,\,0.41i\}$ for $g = 2\pi \times 218~\mathrm{Hz}$ and $\psi_- = \{0.41i,\,0.91\}$ for $g = -2\pi \times 218~\mathrm{Hz}$. 
After the transient ($\sim5~\mathrm{ms}$), the simulated steady-state intensities $\{|\psi_a|^2,\,|\psi_b|^2\} = \{0.83,\,0.17\}$ and $\{0.17,\,0.83\}$ (dashed lines in Figs.~\ref{fig:fig1}(d, e)) agree with these predictions. 
The projection metric $\mathrm{Im}[\langle \psi | H_\mathrm{eff} | \psi \rangle]$ in Figs.~\ref{fig:fig1}(f, g) further verifies collapse onto the attractor branch—the upper band for $g>0$ and the lower band for $g<0$—confirming deterministic eigenmode collapse.

\subsection{Collapse vs.\ Rabi-type Oscillation}

\begin{figure}[h]
    \centering
    \includegraphics[width=0.6\linewidth]{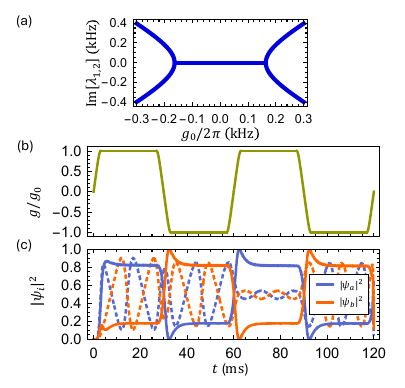}
    \caption{
    (a) Imaginary parts $\mathrm{Im}[\lambda_{1,2}]$ of $H_\mathrm{eff}$ showing $\mathcal{PT}$-symmetric and $\mathcal{PT}$-broken phases separated by the exceptional point $|g_0|=\kappa=2\pi\times163~\mathrm{Hz}$. 
    (b) Temporal modulation of $g$ between $\pm g_0$ for time-dependent eigenmode steering. 
    (c) Simulated $|\psi_i|^2$ for $g_0=2\pi\times218~\mathrm{Hz}$ (solid, broken phase) and $g_0=2\pi\times91~\mathrm{Hz}$ (dashed, symmetric phase). 
    In the broken phase, the system collapses to fixed-point eigenmodes (as in Fig.~\ref{fig:fig1}); in the symmetric phase it exhibits sustained oscillations.
    }
    \label{fig:fig2}
\end{figure}

Having established eigenmode steering via spatial gain–loss distribution, we now introduce temporal control by modulating the gain–loss coefficient in time.  
As shown in Fig.~\ref{fig:fig2}(b), $g(t)$ alternates periodically between $+g_0$ and $-g_0$ (with $g_0 = 2\pi \times 218~\mathrm{Hz}$), connected by $5~\mathrm{ms}$ linear ramps.  
This creates a sequence of temporal boundaries that steer the system in real time between the two attractor eigenmodes of $H_\mathrm{eff}$.  
The simulated intensities $|\psi_{a,b}|^2$ in Fig.~\ref{fig:fig2}(c) capture this evolution: following each reversal of $g(t)$, the system transitions from $\psi_+$ to $\psi_-$ (or vice versa), completing a full cycle of time-dependent eigenmode steering.  
When the modulation depth is reduced to $g_0 = 2\pi \times 91~\mathrm{Hz}$, the convergence after each switch is incomplete, leading to sustained, Rabi-like oscillations between the two modes.

This behavior originates from the $\mathcal{PT}$ symmetry of $H_\mathrm{eff}$, which satisfies $PT H_\mathrm{eff} = H_\mathrm{eff} PT$ with
\[
P = 
\begin{pmatrix}
0 & 1\\[3pt]
1 & 0
\end{pmatrix},
\qquad
T: \text{complex conjugation.}
\]
The symmetry divides parameter space into two regimes separated by an exceptional point (EP) at $|g_0| = \kappa = 2\pi \times 163~\mathrm{Hz}$.
As shown in Fig.~\ref{fig:fig2}(a), the eigenvalues are complex conjugates in the $\mathcal{PT}$-broken phase ($|g_0|>\kappa$) and purely real in the $\mathcal{PT}$-symmetric phase ($|g_0|<\kappa$).
Equation~(\ref{eq:ratio_dynamics}) shows that the modal amplitude ratio is driven by the eigenvalue separation:
\[
\frac{d}{dt}\ln\!\left(\frac{a_1}{a_2}\right) = -\,i(\lambda_1 - \lambda_2).
\]
When $\mathrm{Im}[\lambda_1-\lambda_2]>0$ (broken phase), $|a_1/a_2|$ grows exponentially, yielding deterministic collapse to the dominant eigenmode. 
The imaginary gap $\mathrm{Im}[\lambda_1-\lambda_2]$ sets the characteristic convergence time $\tau_c \!\sim\! 1/\mathrm{Im}[\lambda_1-\lambda_2]$, providing a universal indicator of modal stability and collapse rate.
When $\mathrm{Im}[\lambda_1-\lambda_2]=0$ (symmetric phase), both modes remain balanced, producing sustained Rabi-like oscillations.

Unlike conventional $\mathcal{PT}$-symmetric platforms where total energy diverges or decays, the present cross-coupled, energy-conserving framework yields sustainable, bounded dynamics in both regimes and enables smooth cascading across successive stages of time-dependent eigenmode steering. 
Moreover, the eigenvalue spectrum of $H_\mathrm{eff}$ directly identifies attractor and repulsor modes without a separate Jacobian analysis, offering a global, predictive view of stability.

\subsection{Trimer Model: Cyclic Eigenmode Steering and Energy Routing}

We now extend from two to three coupled resonators to demonstrate how spatiotemporal control routes energy across multiple targets. 
This trimer is the minimal network supporting sequential redistribution and symmetry-mediated transitions. 
Its dynamics obey
\begin{equation}
    i\frac{d}{dt}
    \begin{pmatrix}
        \psi_a \\[3pt]
        \psi_b \\[3pt]
        \psi_c
    \end{pmatrix}
    =
    H
    \begin{pmatrix}
        \psi_a \\[3pt]
        \psi_b \\[3pt]
        \psi_c
    \end{pmatrix},
    \label{eq:trimer_dynamics}
\end{equation}
where $H$ contains Hermitian coupling and nonlinear non-Hermitian feedback. 
For clarity, we consider three identical resonators with identical bridges. 
We decompose $H = H_3 + H_\mathrm{NL}$, where
\begin{equation}
    H_3 = \frac{1}{2}
    \begin{pmatrix}
        0 & -\kappa/2 & -\kappa/2 \\[4pt]
        -\kappa/2 & 0 & -\kappa/2 \\[4pt]
        -\kappa/2 & -\kappa/2 & 0
    \end{pmatrix},
    \label{eq:H3}
\end{equation}
is the symmetric Hermitian coupling with the common diagonal absorbed into the carrier, consistent with the envelope description.
The nonlinear non-Hermitian term, implementing cross-coupled gain–loss feedback, is diagonal:
\begin{equation}
\begin{aligned}
    H_\mathrm{NL} = &\frac{i}{2E}\,
    \mathrm{diag}\!\bigl(
        g_{ab}|\psi_b|^2 + g_{ac}|\psi_c|^2,\;\\
        &g_{bc}|\psi_c|^2 + g_{ba}|\psi_a|^2,\;
        g_{ca}|\psi_a|^2 + g_{cb}|\psi_b|^2
    \bigr),
    \label{eq:HNL}
\end{aligned}
\end{equation}
where $E = |\psi_a|^2 + |\psi_b|^2 + |\psi_c|^2$ and $g_{ij} = g_i - g_j$ are cross-site gain–loss contrasts derived from the site-based \emph{cross-coupled gain--loss coefficients} $g_i$ assigned to $i\in\{a,b,c\}$. 
Only differences $g_{ij}$ carry physical meaning; this construction ensures total energy conservation (see Supplementary Information).

Following the dimer, we introduce an effective Hamiltonian for modal dynamics:
\begin{equation}
    H_\mathrm{eff} = \frac{1}{2}
    \begin{pmatrix}
        i g_a & -\kappa/2 & -\kappa/2 \\[4pt]
        -\kappa/2 & i g_b & -\kappa/2 \\[4pt]
        -\kappa/2 & -\kappa/2 & i g_c
    \end{pmatrix}.
    \label{eq:Heff_trimer}
\end{equation}
The full Hamiltonian takes $H = H_\mathrm{eff} - i G_3 I_3$, with 
\[
G_3 = \frac{1}{2E}\bigl(g_a |\psi_a|^2 + g_b |\psi_b|^2 + g_c |\psi_c|^2\bigr),
\]
and $I_3$ the identity. 
Here, $H_\mathrm{eff}$ sets the modal structure, while $G_3$ regulates amplitude to conserve energy. 
As in the dimer, for any two eigenmodes $\psi_i,\psi_j$,
\[
\frac{a_i}{a_j} \propto e^{-i(\lambda_i - \lambda_j)t},
\]
so the eigenmode associated with the eigenvalue having the largest imaginary part dominates (given nonzero initial projection). 
Thus the direction of energy routing and the steady-state distribution are programmed by the spatiotemporal schedule of $g_a$, $g_b$, and $g_c$.

\begin{figure}[h]
    \centering
    \includegraphics[width=0.6\linewidth]{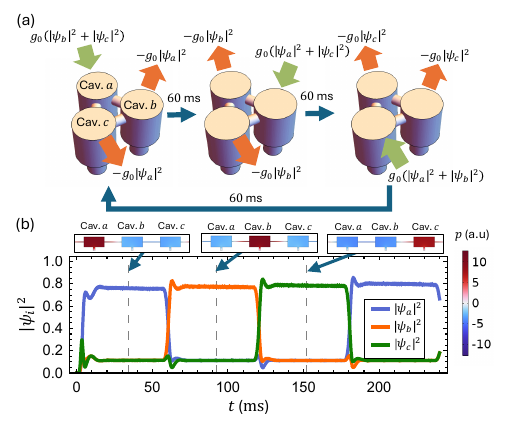}
    \caption{
    (a) Schematic of energy routing in the nonlinear acoustic trimer of Eq.~(\ref{eq:trimer_dynamics}). 
    Cyclic spatiotemporal modulation applies a relative gain $g_0$ sequentially for 60~ms: 
    $(g_a, g_b, g_c) = (g_0, 0, 0) \rightarrow (0, g_0, 0) \rightarrow (0, 0, g_0) \rightarrow \cdots$, 
    steering modal dominance $a \!\rightarrow\! b \!\rightarrow\! c \!\rightarrow\! \cdots$. 
    (b) Simulated $|\psi_{a,b,c}|^2$ showing periodic convergence to distinct fixed points during each step. 
    Representative pressure-field distributions are shown in inset for each interval.
    }
    \label{fig:fig3}
\end{figure}

Figure~\ref{fig:fig3} demonstrates programmable cyclic \emph{eigenmode steering} and energy routing.  
As shown in Fig.~\ref{fig:fig3}(a), the site coefficients $\{g_a,g_b,g_c\}$ are modulated in 60-ms intervals to realize the sequence 
$a\!\rightarrow\!b\!\rightarrow\!c\!\rightarrow\!a$.  
During each interval, one cavity is activated with gain ($g_i=g_0>0$), while the others remain neutral ($g_j=g_k=0$).  
The arrows and accompanying formulas in Fig.~\ref{fig:fig3}(a) follow directly from the nonlinear feedback term in Eq.~(\ref{eq:HNL}),  
where the local contribution for each cavity takes the form  
$\tfrac{i}{2E}\!\left(g_{ij}|\psi_j|^2+g_{ik}|\psi_k|^2\right)$ with $g_{ij}=g_i-g_j$.  
A positive imaginary value (green arrow) represents gain fed by neighbouring intensities, while a negative value (orange arrow) denotes the corresponding loss channel.  
For example, when $(g_a,g_b,g_c)=(g_0,0,0)$, cavity~$a$ acquires gain $\propto |\psi_b|^2+|\psi_c|^2$, whereas $b$ and $c$ experience equal losses $\propto |\psi_a|^2$.  
As the active gain site switches sequentially to $b$ and then to $c$, the dominant eigenmode of $H_\mathrm{eff}$ shifts accordingly, guiding energy around a closed loop through a time-dependent attractor landscape.
The simulated intensities in Fig.~\ref{fig:fig3}(b) confirm precise cyclic routing using the same parameters as the dimer ($g_0 = 2\pi\times218~\mathrm{Hz}$, $\kappa = 2\pi\times163~\mathrm{Hz}$). 
For instance, $(g_a,g_b,g_c)=(g_0,0,0)$ yields the dominating eigenmode $\psi_1=\{-1.47-2.03i,\,1,\,1\}$ (eigenvector of Eq. (\ref{eq:Heff_trimer}) having eigenvalue of the largest imaginary part) with maximal amplitude in cavity~$a$. 
The observed cyclic switching of modal dominance demonstrates deterministic routing via programmed spatiotemporal gain–loss modulation.

\subsection{Overcoming Symmetry-Forbidden Transitions via Spatiotemporal Perturbation}

\begin{figure}[t]
    \centering
    \includegraphics[width=0.6\linewidth]{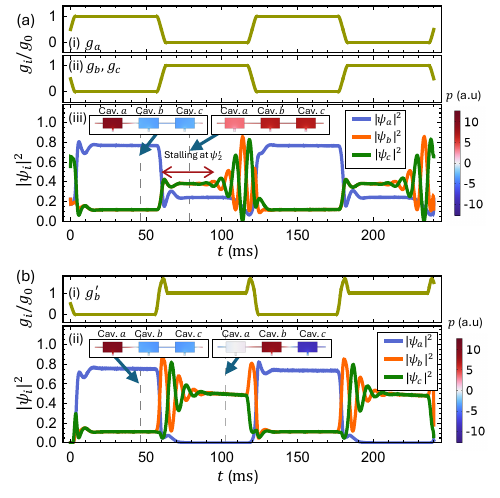}
    \caption{
    (a) Spatiotemporal profile for redistributing total energy to cavities $b$ and $c$. 
    (i) Modulation schemes for $g_a$, (ii) for $g_b, g_c$ with symmetric modulation ($g_b = g_c$). 
    (iii) Simulated $|\psi_{a,b,c}|^2$. 
    Energy concentrates in cavity $a$ during the first 60~ms (matching Fig.~\ref{fig:fig3}(b)) but fails to redistribute to $b$ and $c$ during $60 < t < 120$~ms, unable to reach the target eigenmode $\psi_1' = \{0, 1, -1\}$. 
    (b) Successful redistribution using a \emph{spatiotemporal perturbation} applied as a transient modification $g_b'(t)$ in (i). 
    (ii) Simulated $|\psi_{a,b,c}|^2$ show that this perturbation enables the previously symmetry-forbidden transition, with convergence to $\psi_1'$ at $t \simeq 100$~ms. 
    Pressure field patterns are sampled and shown for each interval.
    }
    \label{fig:fig4}
\end{figure}

We now demonstrate assisted \emph{eigenmode steering} for a transition that is \emph{symmetry-forbidden} under ideal conditions. 
We switch the site schedule from a single-cavity preference 
$(g_a, g_b, g_c) = (g_0, 0, 0)$ to a dual-cavity configuration $(g_a, g_b, g_c) = (0, g_0, g_0)$, 
aiming to redistribute energy from resonator~$a$ into equal amplitudes in resonators~$b$ and~$c$ (Fig.~\ref{fig:fig4}(a)(i, ii)). 
The target attractor eigenmode of the dual-cavity configuration is 
$\psi_1' = \{0, 1, -1\}$ (largest imaginary part of eigenvalue), which suppresses cavity~$a$ while establishing equal and opposite phases in $b$ and $c$.

Full-wave simulations in Fig.~\ref{fig:fig4}(a)(iii) show that this transition does not occur spontaneously. 
After the gain reconfiguration at $t = 60~\mathrm{ms}$, the system becomes trapped in an intermediate state 
$\psi_2' = \{0.47 - 0.64i,\, 1,\, 1\}$ instead of collapsing to the antisymmetric target (see Supplementary Information). 
Energy partially transfers from $a$ (blue) to $b$ and $c$ (orange/green), but cavity~$a$ retains significant amplitude.
This trapping arises from a parity mismatch: 
the initial eigenmode $\psi_1$ has $b$ and $c$ in phase, whereas the target $\psi_1'$ requires them out of phase. 
Because $(0, g_0, g_0)$ is parity-symmetric, it cannot couple even- and odd-parity modes; the transition is therefore symmetry-forbidden in the ideal, noiseless system.

To enable and accelerate the transition, we introduce a controlled \emph{spatiotemporal perturbation}—a transient modification to $g_b(t)$ (leaving $g_a(t)$ unperturbed), as in Fig.~\ref{fig:fig4}(b)(i). 
This momentarily breaks the $b$--$c$ symmetry at the switching points, lifting the parity constraint and directing convergence to the target eigenmode. 
Figure~\ref{fig:fig4}(b)(ii) shows rapid convergence to $\psi_1'$ within $t \approx 100~\mathrm{ms}$, achieving complete redistribution: resonator~$a$ is fully suppressed while $b$ and $c$ are equally amplified with opposite phase. 
Thus temporal modulation governs not only eigenmode selection but also the transition pathway; small, designed spatiotemporal perturbations can dramatically enhance convergence rates by removing symmetry bottlenecks. 
In this framework, the spatial profile defines the target eigenmodes, while temporal perturbations program the kinetics of steering toward them.

\section{Conclusion}

We have established a spatiotemporal gain–loss framework that enables \emph{eigenmode steering} in coupled acoustic resonators.  
The dynamics are governed by a nonlinear Hamiltonian featuring a \emph{cross-coupled gain–loss coefficient}, which enforces total energy conservation while guiding the system toward fixed points corresponding to the eigenvectors of an effective non-Hermitian Hamiltonian.  
Within this framework, the eigenmode associated with the eigenvalue having the largest imaginary part naturally emerges as the attractor, realizing a deterministic form of \emph{eigenmode collapse}.  

Time-dependent modulation of the gain–loss coefficient reveals two distinct regimes divided by the exceptional point:  
the $\mathcal{PT}$-broken phase, exhibiting stable eigenmode convergence, and the $\mathcal{PT}$-symmetric phase, characterized by sustained Rabi-like oscillations.  
Extending the concept to a trimer system demonstrates programmable and cyclic \emph{eigenmode steering} across multiple resonators.  
Moreover, introducing controlled \emph{spatiotemporal perturbations} enables transitions that are otherwise \emph{symmetry-forbidden}, providing an additional temporal degree of control that accelerates convergence and lifts symmetry constraints.

Validated by full-wave simulations of coupled Helmholtz resonators, this framework unifies nonlinear feedback, $\mathcal{PT}$ symmetry, and time-varying control into a single platform for sustainable and reconfigurable acoustic manipulation.  
It opens pathways toward adaptive noise control, programmable metamaterials, and analog information processing, and offers broader insight into how spatiotemporal gain–loss engineering can shape wave dynamics in non-Hermitian and energy-conserving systems.

\medskip
\noindent\textbf{Acknowledgements}
The authors acknowledge financial support from the EPSRC via the META4D Programme Grant (No. EP/Y015673/1). 

\bibliography{bib}% common bib file

\end{document}